\def\year{2019}\relax
\documentclass[letterpaper]{article} 
\usepackage{aaai19}  
\usepackage{times}  
\usepackage{helvet}  
\usepackage{courier}  
\usepackage{url}  
\usepackage{graphicx}  
\newcommand{\csx}{CiteSeerX}

\usepackage{amsmath}
\usepackage{algorithm}
\usepackage{algpseudocode}
\usepackage{amsfonts}
\usepackage{threeparttable}
\usepackage{array}
\usepackage{multirow}
\usepackage{comment}

\begin{document}
%
\title{Cleaning Noisy and Heterogeneous Metadata for Record Linking Across \\
Scholarly Big Datasets}
\author{%
 \parbox{\textwidth}{\centering
{Athar Sefid}\textsuperscript{1},
{Jian Wu}\textsuperscript{3},
{Allen C. Ge}\textsuperscript{2},
{Jing Zhao}\textsuperscript{2},
{Lu Liu}\textsuperscript{2},\\
{Cornelia Caragea}\textsuperscript{4},
{Prasenjit Mitra}\textsuperscript{2},
{C. Lee Giles}\textsuperscript{1,2}}  \\ 
{\textsuperscript{1}\large Computer Science and Engineering, Pennsylvania State University, University Park, PA, 16801}\\
{\textsuperscript{2}\large Information Sciences and Technology, Pennsylvania State University, University Park, PA, 16801}\\
{\textsuperscript{3}\large Computer Science, Old Dominion University, Norfolk, VA, 23529}\\
{\textsuperscript{4}\large Computer Science, University of Illinois at Chicago, Chicago, IL, 60607} \\
{\tt azs5955@psu.edu},
{\tt jwu@cs.odu.edu}
}
\maketitle
\begin{abstract}
Automatically extracted metadata from scholarly documents in PDF formats is usually noisy and heterogeneous, often containing incomplete fields and erroneous values. One common way of cleaning  metadata is to use a bibliographic reference dataset. The challenge is to match records between corpora with high precision. The existing solution which is based on information retrieval and string similarity on titles works well only if the titles are cleaned. We introduce a system designed to match scholarly document entities with noisy metadata against a reference dataset. The blocking function uses the classic BM25 algorithm to find the matching candidates from the reference data that has been indexed by ElasticSearch. The core components use supervised methods which combine features extracted from all available metadata fields. The system also leverages available citation information to match entities. The combination of metadata and citation achieves high accuracy that significantly outperforms the baseline method on the same test dataset. We apply this system to match the database of CiteSeerX against Web of Science, PubMed, and DBLP. This method will be deployed in the CiteSeerX system to clean metadata and link records to other scholarly big datasets. 
\end{abstract}

\section{Introduction}
\noindent Since the advent of Scholarly Big Data (SBD)\cite{giles2013sbd}, there has been a growing interest in topics related to this big data instance, such as scholarly article discovery \cite{wesleysmith2016wwwbigscholar}, semantic analysis \cite{al2017machine}, recommendation systems \cite{huang2015aaai-citationrecommendation}, citation prediction \cite{liu2017predictive}, scalability improvement \cite{kim2017scaling}, and Science of Science \cite{chen2018aaai}. Major SBD datasets include the Microsoft Academic Graph (MAG), \csx\ \cite{giles1998jcdl,wu2014iaai}, DBLP, Web of Science (WoS), and Medline, among which MAG and \csx\ are the only two freely available large scale datasets which offer citation graphs. Academic search engines such as Microsoft Academic and \csx\ obtain raw PDF files by actively crawling the Web. These PDF documents are then classified into academic and non-academic documents. Metadata, citations, and other types of content are then extracted from these documents. Different from submission-based datasets such as the WoS, a large fraction of documents crawled are pre-prints and manuscripts, which do not necessarily contain unique identifiers, e.g., DOIs. Metadata from these documents may also differ from their official published versions. Also, errors may occur when text is extracted from PDFs, and when metadata is parsed. As a result, metadata extracted is likely to be incomplete and erroneous. This metadata is also heterogeneous since the documents were written by authors using different conventions and templates. This noisy information can propagate through to data analytics and aggregations that can then distort research, making cleaning it a necessity. One common approach is to link document entities to corresponding entities in a cleaned dataset (reference dataset) and then use its records to replace the automatically extracted records. 

The biggest challenge of this approach is to find the correct matching entity in the reference dataset. Though different from traditional machine learning (ML) tasks in which there is a trade-off between precision and recalls, the entity matching task must be accomplished with \emph{high precision}. This is because when the dataset is large, a small fraction of false positives may lead to a large number of ``false corrections''. The problem is even more challenging because there is usually no prior knowledge of which fields are noisy. 

Our contributions include:
(1) developing a ML-based paper entity matching framework which includes both header and citation information of available scholarly documents; (2) applying the system on CiteSeerX, WoS, DBLP, and PubMed to find  overlap between these digital library databases which is used to clean \csx\ data. Plus this generates a more accurate citation graph data and links records which can enrich the content of individual document. 
\section{Related Work}

There are in general three types of methods in entity matching across bibliographic databases.

{\bf Information Retrieval-based}: This method searches one or multiple attributes of an entity in the target corpus against the index of the reference corpus and rank the candidates using a similarity metric. This approach matches \csx\ with DBLP  \cite{caragea2014ecir}. The reference dataset (DBLP) was indexed by Apache Solr. Metadata from the noisy dataset (\csx) were used to query corresponding fields. Candidates were selected based on similarity scores. It was found that using 3-grams of titles and Jaccard similarity with a threshold of $0.7$ achieves the best F1-measure of $77\%$. Because of the relatively low precision, the approach cannot be directly used in cleaning  \csx\ data. 

{\bf Machine Learning-based}: These methods have been used in entity matching of user entities in online social networks \cite{peled2013entitymatching}. The problem is to match user profiles on Facebook and Xing.  This work applies pairwise comparison to whole dataset ($\approx15,000$ records) without applying any blocking function to reduce search spaces, so this method cannot be scaled up to large digital libraries containing tens of millions of records. 

{\bf Topical-based}: This method is used to resolve and match entities that are represented by free text, e.g., Wiki articles. The challenge is that different sources may use different languages or terminologies to describe the same topic. A probabilistic model was proposed to integrate the topic extraction and matching into a unified model \cite{yang2015sigkdd}. As we don't have access to the full text of the reference datasets, this method is not applicable to our problem.  

\section{Models}
\subsection{Entity Representation}
Throughout below, we refer to a scholarly paper with full information as a \emph{paper entity}. We denote the target corpus which contains noisy data as $T$. This contains $n$ paper entities $t_i,1\leq i\leq n$, and a reference corpus $R$ which contains reference data and $m$ paper entities $r_j,1\leq j\leq m$.
Each entity can be represented by a number of attributes. 
Our goal is to find a set $M$, 
$$M=\left\{(t,r);t=r,t\in T, r\in R\right\}.$$

An officially published paper usually is assigned a unique identifier, i.e. DOI. A journal article can also be identified by the journal name, volume, issue number, and the starting page number. However, for a large fraction of open access scholarly papers crawled from the Web, such information is usually not available. Empirically, a paper entity can be uniquely identified by four header fields, (title, authors, year, venue), 
in which the venue is a conference or a journal name. 
A citation record parsed from a citation string usually contains the above four fields. So, matching a single citation record can be done in the same manner as matching a paper entity. The abstract is usually a paragraph of text that may contain non-ASCII characters with different encodings, but normalizing abstracts and calculating simhash values takes heavy overhead, so in this work, we use abstracts without normalization. Due to a lack of general venue disambiguator, venue information is not incorporated as matching features. We will show that even without it, the application still achieves high performance.

Paper entity linking can be formalized as a binary classification problem in which the classifier decides whether a candidate pair is a real match or not. Because many digital library datasets do not have citation information, we consider two separate models, one with only header information called Header Matching Model (HMM), and the other with only citation information called Citation Matching Model (CMM). The combination of them is called the Integrated Matching Model (IMM). There is a separate model to evaluate title quality called Title Evaluation Model (TEM).

\subsection{Header Matching Model (HMM)}
\label{hmm}
{\bf HMM} is a supervised machine learning model to classify candidate pairs using information from the paper header. 
We first index all document metadata in 
the reference corpus using an open source search platform. 
We use ElasticSearch (ES) because of its relatively low setup overhead and scalability. The default settings are applied for our experiments. The indexed metadata contains header fields including titles, authors, abstracts, and years. 
For each paper in the target corpus, we query the title against the index, if it contains a minimum of 20 characters. Otherwise, the first author's last name and publication year are used in queries. If the year is not available, only the first author's last name is used. For each field, the query string is segmented into unigrams connected by ``OR''. The ranking scores are calculated using the Okapi BM25 algorithm \cite{robertson2004}. The query algorithm is shown in Algorithm~\ref{query}.
 
For each target paper, the top 10 papers from the reference set are retrieved and 10 candidates are formed as matching pairs. The features of each pair include a list of similarities calculated using the header metadata. 
\begin{enumerate}
    \item Title similarity represented by Levenshtein distance of simhashes of normalized titles \cite{charika2002stoc_simhash}:
    \begin{align}
    \label{simmash}
    \mbox{Sim}_L(\mbox{title}_{t},\mbox{title}_{r})=\mbox{lev}_{a,b}\left(|a|,|b|\right)\\
    a=\mbox{Simhash}_{16}(\mbox{Norm}(\mbox{title}_t))\\b=\mbox{Simhash}_{16}(\mbox{Norm}(\mbox{title}_r))
    \end{align}
    
    in which a simhash string contains 16 alphanumeric characters. The titles are normalized so that (1) all letters are lowercased; (2) diacritics are removed, e.g., ``\'{a}'' is converted to ``a''; (3) consecutive spaces are collapsed; (4) punctuation marks are trimmed off; (5) single characters ``s'' and ``t'' are removed because they are mostly resulted from removing apostrophe from possessives or abbreviations such as ``can't''.
    \item Abstract similarity $\mbox{Sim}_L(\mbox{abstract}_t,\mbox{abstract}_r)$
    represented by Levenshtein distance of simhashes of abstracts, calculated in a similar way as Equations~(1)--(3) without normalization.
     \item Jaccard similarities between normalized titles and original abstracts. For example, 
     \begin{equation}
     \label{bowmodel}
     \mbox{Sim}_J(\mbox{title}_t,\mbox{title}_r)=\frac{|W_t\cap W_r|}{|W_t\cup W_r|}
     \end{equation}
     in which $W_t$ and $W_r$ represent the token set of the title of the target and the reference paper, respectively. 
         \item The absolute difference of the years.
	\item The {\bf first and the last author's full name similarity}. Author similarities are measured in multiple metrics. An author's full name similarity is represented by a three digit binary $lmf$, representing whether the last name, the middle initial, and the first initial matches, respectively. If a certain name component is missing or it does not match, the binary is set to $0$. The decimal value of a binary is used as the full name similarity index. Author names are also normalized before comparison. Diacritics are removed and letters are lowercased. Prefixes, e.g., Prof., and suffixes, e.g., ``II'', and their variants are removed. For example, if first authors are ``Jane C. Huck'' and ``J. Huck'', the binary is 101, which equals to 5 in decimal. 
    \item  The {\bf last name similarities of the first and the last author.} The last name similarity is computed in this way
    \begin{equation}
    \label{namesim}
    \small{
    \mbox{Sim}(\mbox{N}_{t},\mbox{N}_{r})=\left\{\begin{array}{l@{:}l}
    0\ &\ \mbox{N}_{t}\neq\mbox{NULL} \land \mbox{N}_{r}\neq\mbox{NULL} \land \mbox{N}_{t}\neq\mbox{N}_r \\
    1\ &\ \mbox{N}_{t}=\mbox{NULL} \lor \mbox{N}_{r}=\mbox{NULL} \\
    2\ &\ \mbox{N}_{t}\neq\mbox{NULL} \land \mbox{N}_{r}\neq\mbox{NULL} \land \mbox{N}_{t}=\mbox{N}_r\\
    
    \end{array} \right.
    }
    \end{equation}
    in which N stands for the last name and NULL means the value is not available. 
   \item {\bf All authors' last name Jaccard similarity}    
   \begin{equation}
   \small{
   \label{allauthor}
   \mbox{Sim}_J\left(L_{t},L_{r}\right)=\frac{|L_t\cap L_r|}{|L_t\cup L_r|}
   }
   \end{equation}
   in which $L_t$ and $L_r$ stands for the set of last names in the target and the reference corpora, respectively. 
    \end{enumerate}

The pseudocodes of the HMM is in Algorithm~\ref{hmmalg}. 

\begin{algorithm}[htb]
	\caption{Query Builder}
	\label{query}\small
	\begin{algorithmic}[1]
		\Function{Query}{$ title, lastName, year $}
        	\If {title $\neq$ Null and title.length $>$ 20}
                  	\State query $\gets$ title
                  \ElsIf { lastName $\neq$ Null and year $\neq$ Null}
                  	\State query $\gets$ lastName and year
                  \ElsIf{ lastName $\neq$ Null }
                  	\State query $\gets$ lastName
                  \EndIf
        	\State return query            
		\EndFunction
	\end{algorithmic}
\end{algorithm}

\begin{algorithm}[htb]
	\caption{Header Matching Model}
	\label{hmmalg}\small
	\begin{algorithmic}[1]
		\Function{HMM}{$ $}
        	\State T $\gets$ target corpus
            \State R $\gets$ reference corpus
        	\State $index_R$ $\gets$ index of reference corpus
            \State matchList $\gets$\O
			\For{$t \in T$}
				  \State Q $\gets$ Query($t$.title, $t$.firstLastName, $t$.year)
        		  \State Candidates $\gets$ query Q to $index_R$
                  \For{c $\in$ Candidates}
                  	  \State prediction$\gets$Model.predict($t$,c)
               		  \If{prediction=1}
                      	\State   matchList.add $(t,c)$  
                        \State break
                      \EndIf
                  \EndFor 
        	\EndFor 
		\EndFunction
	\end{algorithmic}
\end{algorithm}

\begin{algorithm}[!h]
	\caption{Citation Matching Model}
	\label{citalgo}\small
	\begin{algorithmic}[1]
		\Function{CMM}{$ $}
            \State T $\gets$ target corpus
            \State R $\gets$ reference corpus
            \State matchList $\gets$\O
        	\State $CitationIndex_R$ $\gets$ citations index of reference corpus
			\For{$t \in T$}
            	  \State $t\_citations$ $\gets$ citations of $t$. 
                  \For{$ tc_i \in t\_citations$  }
                    \State Q $\gets$ Query ($tc_i$.title,$tc_i$.firstLastName, $tc_i$.year)
        		  	\State results $\gets$ query Q to $CitationIndex_R$
                  	\For{$rc_j$ $\in$ results}
                  	  \State prediction$\gets$Model.predict ($tc_i$, $rc_j$)
               		  \If{prediction=1}\Comment{citations match}
                      	\State $r_k$ $\gets$ paper that cites $rc_j$
                        \State $r$.title $\gets$ Simhash ($r_k$.title)
                        \State $t$.title $\gets$ Simhash ($t$.title)
                        \State $title\_dist$ = lev ($t$.title, $r$.title)
                        \If{$title\_dist < \theta_{title}$}
							\State matchList.add ($t$, $r_k$)
                            \State break
                        \EndIf
                        
                        \State $BoW_{r}$ $\gets$ BoW ($r_k$.referenceTitles) 
                        \State $BoW_{t}$ $\gets$ BoW ($t$.referenceTitles)
                         \State $sim$ $\gets$ Jaccard($BoW_{t_i}$,  $BoW_{r_i}$)
                         \If{$sim > \theta_{ref}$}
                            \State matchList.add ($t$, $r_k$)
						    \State break
                            
                        \EndIf
                      \EndIf
                  \EndFor
                  \EndFor
        	\EndFor 
		\EndFunction
	\end{algorithmic}
\end{algorithm}

\subsection{Citation Matching Model (CMM)}
{\bf CMM} matches paper entities by citations. The paper entity matching problem can benefit from this model when the header metadata is noisy but the references are available. Similar to HMM, a prerequisite is to index  all citations in the reference corpus. On average, one paper contains about 20 citations , so the citation index is usually much larger than the document index.

Given a target paper $t$, its citation records $tc_i$ are retrieved from the database. We attempt to find the matching record $rc_j$ in the reference corpus using the query builder in Algorithm \ref{query}. Retrieved citations and $rc_i$ are matched by the HMM (Algorithm \ref{hmmalg}). Citations do not contain abstracts so relevant features are not used. 
Assuming such $j=1$ exists (if not, then no matching entity is found) and $rc_1$ is cited by $r_1$, the next step is to compare $r_1$ with $t$. The CMM uses both the \emph{paper title} and the \emph{citation titles} (Algorithm~\ref{citalgo}). First, the title similarities are calculated using Equations (1)--(3). If this distance is less than a threshold of $\theta_{title}$, $r_1$ is believed to be the matching entity of $t$. Otherwise, CMM extracts the tokens from all the reference titles of $t$, denoted by $BoW_{t}$, and tokens from all the reference titles of $r_1$, denoted by $BoW_{r}$. The judgment is made by comparing the Jaccard similarity between $BoW_{t}$ and $BoW_{r}$. If the similarity is greater than a threshold $\theta_{ref}$, then $(t,r_1)$ is determined as a matching pair. Otherwise, the algorithm continues to examine the next paper that shares citation $rc_1$ with $t$. If no papers citing $rc_1$ is found to be a match for $t$, CMM  continues and attempts to find the matching record of $tc_2$. 

\subsection{Title Evaluation Model (TEM)}
{\bf TEM} is a light-weight supervised learning model designed to provide a quantitative evaluation of the title quality. The input is a title string, and the output is a probability $\theta$ of how likely the input string looks like a paper title. The title quality is determined to be high if $\theta$ is greater than a threshold $\theta_{tq}$. The TEM exploits the features in Table~\ref{temfeatures} extracted from the original title string. The TEM is trained on a sample of 8200 title strings containing 6270 high-quality titles and 1930 titles with low quality. 
Titles are labeled to be of low quality if (1) They are NULL; (2) They have many non-ASCII characters; (3) They include evidently irrelevant information such as authors; (4) They are not in English. 

Four supervised models, including Logistic Regression (LR), Support Vector Machine (SVM), Na\"{i}ve Bayes (NB), and Random Forests (RF), are trained. The LR model achieves the best 10-fold cross-validated F1-score of $0.999$, which we adopted. 

\begin{table}[t]
\begin{threeparttable}
\centering\footnotesize
\caption{\label{temfeatures}\small Features used to train TEM.}
\begin{tabular}{m{8cm}}
\hline\hline
{\bf Character-level features}\\ \hline
\#ASCII characters\quad \#non-ASCII characters\quad \#white spaces\\
\#punctuation marks\quad \#consecutive punctuation marks\ \ \ \#digits\\
Type of the first/last character (punctuation, digit, or letter)\\
\hline
{\bf Word-level features}\\ \hline
\#$\max{(\mbox{DF}(w))},w\notin\mathbb{S}$\tnote{1} \quad \#$\min{(\mbox{DF}(w))},w\notin\mathbb{S}$ \\
\#$\mbox{median}(\mbox{DF}(w)),w\notin\mathbb{S}$\quad \#words\\
\#Appearance of one of the tokens in the controlled list\tnote{2}: \{Abstract, List, Acknowledgments, Notices, Content, Accepted, Authors, References, Acknowledgments, Null, Chapter, Discussions, Summary\} \\
\hline
\end{tabular}
\begin{tablenotes}
\item [1] DF: document frequency, calculated on all DBLP titles. $\mathbb{S}$ is a set of stopwords adopted from Apache Solr. 
\item [2] The value is set to 1 if the string contains at least one exact match to the controlled list.
\end{tablenotes}
\end{threeparttable}
\end{table}

\begin{figure}[htb]
\centering
       \includegraphics[width=0.5\textwidth]{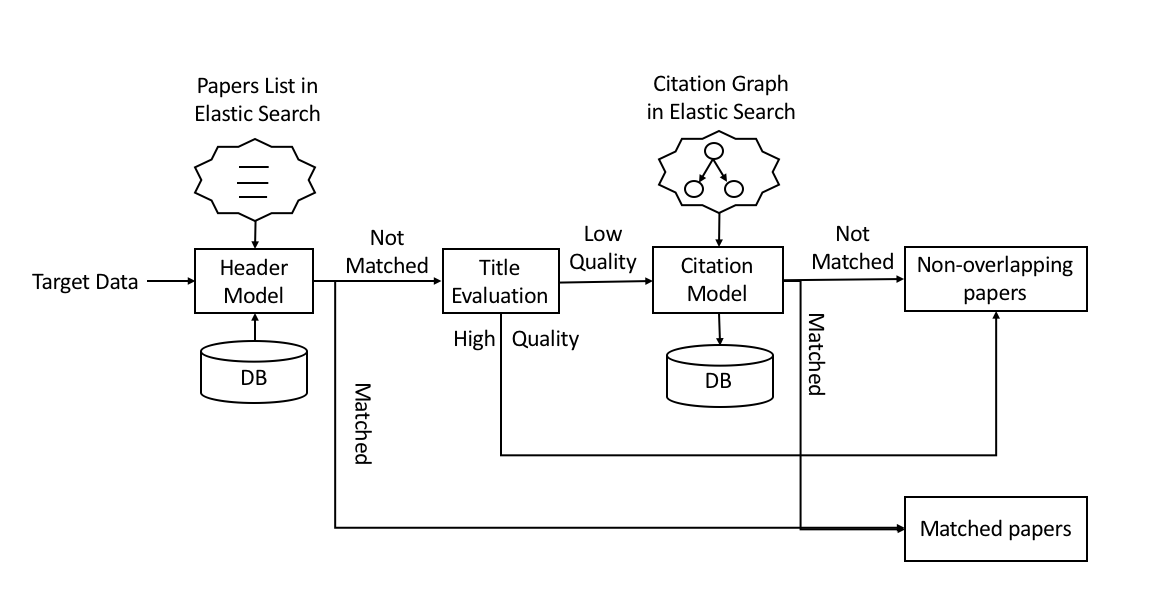}
\caption{\label{flowchart} The pipeline of IMM.}
\end{figure}

{\bf IMM} integrates HMM, CMM, and TEM (Figure~\ref{flowchart}).  If HMM is able to find the match of a paper entity, the process continues to the next paper. Otherwise, the paper title quality is evaluated by TEM. If the title quality is high  ($\theta\geq\theta_{tq}$), it is likely that there is not a matching entity existing in the reference corpus, otherwise ($\theta<\theta_{tq}$), the matching entity is not found due to the poor title quality. In this situation, we use citation information to match papers.

\section{Experiments \label{Experiments}}
\subsection{Data}
The target data is the \csx\ database with about 9 million scholarly papers. The header metadata is extracted by GROBID. References are extracted and parsed by ParsCit \cite{councill2008parscit}. 
The reference corpora are described below.

{\bf WoS} is a digital library dataset spanning 230+ academic disciplines with citation indexing. WoS indexing coverage is from 1900 to 2015 with over 20,000 journals, books, and conference proceedings. There are about 45 million WoS papers and 906 million citation records in this corpus. 

{\bf DBLP} is a bibliographic dataset covering more than 5,000 conferences and 1,500 journals in computer science. We use the version published in March, 2017 with about 4 million documents. This dataset does not contain citations. 

{\bf Medline} is the premier bibliographic dataset released by NCBI with about 24 million academic papers in the area of biomedicine published since 1966. The dataset does not contain citation information. 

The {\bf IEEE} corpus is a \emph{subset} of the IEEE Xplore database, containing about 2 million bibliographic records downloaded from IEEE FTP sites. It does not contain citations. 

\subsection{Ground Truth Labeling } \label{gtsection}

The procedure for labeling process comprises three steps. First, for each paper in the \csx\ sample set, a candidate set of 10 papers is retrieved from the reference index using the same manner described in Algorithm~\ref{query}. Then, to determine the true matches out of candidate papers, other metadata of the papers including authors, abstract, year, venue, keywords, and the number of pages were visually inspected independently by two graduate students. Finally, if it was not possible to decide based on papers profiles, actual PDF files were used side by side to make final decisions.  We generated the following ground truth datasets: 

{\bf \csx-IEEE} This dataset, adopted from Wu et al. \shortcite{wu2017supervised}, is built based on 1000 \csx\ papers with 51 true matching pairs found in the IEEE corpus.

{\bf \csx-DBLP}  This dataset, revised based on Caragea et al. \shortcite{caragea2014ecir}, contains 292 matching pairs identified between 1000 \csx\ papers and the DBLP dataset.  

{\bf \csx-WoS} This dataset contains 345 matching papers found in WoS out of 533 \csx\ papers. 

{\bf Combined Sample} The positive sample contains 688 matching pairs. The negative samples are selected using 1845 candidate matching pairs, containing the most similar but unmatched papers.

\begin{table}[htb]
\centering
\footnotesize
\begin{threeparttable}
\caption{\label{modelresults}\small HMM model 10-fold CV results.}
\begin{tabular}{c|c|c|c}
\hline
Model & Precision & Recall & F1-measure  \\ 
\hline
{\bf SVM} & {\bf 0.926} & {\bf 0.937} &  {\bf 0.931} \\
LR & 0.794 & 0.968 & 0.872\\
RF & 0.912 & 0.931 & 0.921  \\
XGBoost & 0.925 & 0.899 & 0.912\\ 
\hline
\end{tabular}
\end{threeparttable}
\end{table}

\subsection{Experiment Setups}
We trained binary classifiers that decide whether a pair of documents from target and reference corpora is a true matching pair. 
Four machine learning models, SVM, LR, RF, and XGBoost are trained on the Combined Sample. Grid search is applied to tune and find the hyper parameters yielding the best results. Precision, Recall, and F1-score values for 10-fold CV are reported in Table~\ref{modelresults}. 

\subsection{Results and Discussion}
In four models, SVM achieves the highest F1-measure. RF has a comparable F1-measure but requires a significantly reduced test time ($<30\%$), so we employ RF for HMM. The information gain (IG) is calculated for each feature indicating that the most informative features are related to titles and the first authors. 
 
To make an even comparison with the method proposed by Caragea et al. \shortcite{caragea2014ecir}, we rerun their experiments on the CiteSeerX-DBLP ground truth with the best parameter settings in which $n=3$ and the Jaccard similarity threshold $\theta_J=0.7$. We then compare the results with our RF model trained on the combination of \csx-WoS and \csx-IEEE datasets. 
The HMM outperforms the IR-based model with 14\% improvement in precision ($100\%$ vs. $86\%$) and a 3\% improvement in the F1 score ($91\%$ vs. $88\%$). 
 
We investigate how the reference Jaccard similarity threshold $\theta_{ref}$ and title Levenshtein distance $\theta_{title}$ affects the performance of CMM (Table \ref{cit2}).
A higher value of $\theta_{ref}$ indicates that two papers need more common citations to be considered as a matching pair. The best F1 score is obtained at $\theta_{ref} = 0.5$ and $\theta_{title} = 0.35$. 

\begin{table}[htb]
\centering
\begin{threeparttable}
\caption{\label{cit2}\small The CMM performance with different $\theta_{ref}$ and $\theta_{title}$.}
 \footnotesize
\begin{tabular}{p{1.4cm}|p{1.1cm}|p{1.1cm}|p{1.1cm}|p{1.1cm}}
\hline
$\theta_{ref}$  & $\theta_{title}$ & Precision & Recall & F1  \\ 
\hline
\multirow{5}{*}{0.40} & 0.15& 0.876 & 0.719 & 0.790 \\
& 0.25 & 0.877& 0.725 & 0.794 \\
& 0.35 & 0.878& \textbf{0.730} & 0.797 \\
& 0.45 & 0.850 & 0.725 & 0.782 \\
\hline
\multirow{5}{*}{0.50} & 0.15& 0.968 & 0.690 & 0.797 \\
& 0.25 & 0.969& 0.714 & 0.822 \\
& 0.35 & 0.965& 0.728 & \textbf{0.830} \\
& 0.45 & 0.927 & 0.725 & 0.814 \\
\hline
\multirow{5}{*}{0.60} & 0.15 & 0.982 & 0.651 & 0.783 \\
& 0.25 & \textbf{0.983} & 0.662 & 0.791 \\
& 0.35 & 0.979 & 0.691 & 0.810 \\
& 0.45 & 0.938 & 0.691 & 0.796 \\
\hline
\multirow{5}{*}{0.70} & 0.15 & 0.955 & 0.609 & 0.743 \\
& 0.25 & 0.955 & 0.620 & 0.752 \\
& 0.35 &  0.953& 0.652 & 0.775 \\
& 0.45 & 0. 912& 0.658 & 0.764 \\
\hline
\end{tabular}
\end{threeparttable}
\end{table}

Table~\ref{final_result} compares HMM, CMM, and IMM based on the CiteSeerX-WoS dataset (because only this dataset contains citations). In the first column, as the threshold $\theta_{tq}$ increases from $0.01$ to $0.2$, the testing corpus encloses more papers with higher quality titles, which results in a better performance of HMM. The CMM alone is getting better with remarkably high precision but poor recalls. The integrated model achieves both high recall and precision. This indicates that (1) CMM tend to be more useful when the title quality is low; (2) The integrated model significantly increases the overall performance, especially for papers with low quality titles. 

One result in Table~\ref{final_result} that is counter-intuitive is the HMM consistently achieves high performance when the title quality is low. 
To answer this question, we trained a RF classifier on papers with low-quality titles only. IG of the new model reveals that the most important features in the absence of good titles are First author features, Jaccard similarity of all authors' last names, and Abstract features, implying that when title quality is low, accurate author information can also provide accurate matches. 

\begin{table*}[h]
\centering
\begin{threeparttable}\footnotesize
\caption{\label{final_result}\small  Comparisons of HMM, CMM, and IMM performances using the CiteSeerX-WoS dataset with different title quality thresholds. $T/s$ stands for testing time in seconds.}
\begin{tabular}{c|c|c|c|c|c|c|c|c|c|c|c|c|c}
\hline
\multirow{2}{*}{$\theta<\theta_{tq}$}  & \multicolumn{1}{c|}{\bf Data}& \multicolumn{4}{c}{\bf HMM} & \multicolumn{4}{|c}{\bf CMM} & \multicolumn{4}{|c}{\bf IMM} \\ 
\cline{3-14}
& \multicolumn{1}{c|}{\bf Portion}  &  P & R  & F1  & $T/s$ &  P & R  & F1 & $T/s$ & P & R  & F1 & $T/s$\\
\hline
$\theta<0.01$ & 16.1 \% & 0.971 & 0.872 & 0.919 & 10 & 1.0 & 0.513 & 0.678 & 12229 & 0.975 & 1.0 & 0.987 & 9082 \\
$\theta<0.02$ & 17.8 \% & 0.973 & 0.857 & 0.911 & 12 & 1.0 & 0.524 & 0.688 & 12761 & 0.977 & 1.0 & 0.988 & 9791 \\
$\theta<0.10$ &  21.20\% & 0.978 & 0.833 & 0.900 & 14 & 1.0 & 0.593 & 0.745 & 13732 & 0.982 & 1.0 & 0.991 & 10206 \\
$\theta<0.20$ &  24.39\% & 0.981 & 0.869 & \textbf{0.922} & 14.5 & 1.0 & 0.59 & 0.742 & 14823& 0.984 & 1.0 & \textbf{0.992} & 11490\\
\hline
\end{tabular}
\end{threeparttable}
\end{table*} 
 
\subsection{Error Analysis}
Although the combination of HMM and CMM achieve superior performance, the recall of CMM alone is poor (Table~\ref{final_result}). This could be due to two reasons: (1) Citation parsing errors. For example, more than 1 million papers in \csx\ contain less than 5 citations;  (2) Null title citations. About $17\%$ of citation records in WoS and $8.4\%$ of citations in \csx\ have null titles.

The citation-based model is slow because (1) the large number of citations (906 million) slows down the search process and (2) the candidate set for each \csx\ citation could be huge for highly-cited papers. The integrated model only applies citation model to papers with low-quality titles to improve recall. 

\subsection{Application and Conclusion}
We applied HMM on \csx\ documents against DBLP, WoS, and Medline. 
The result indicates that \textbf{the current \csx\ dataset includes about 3 million WoS documents, 1.62 million Medline papers, and about 1.35 million DBLP papers}. The matching process by HMM is done in 11 days on a machine with following specifications: 32 logical cores of Intel Xeon CPU E5-2630 v3 @ 2.40GH; 330~GB RAM.   
The result reveals that there is still a large number of papers that \csx\ should index. The unmatched document metadata can aid \csx\ crawler to find relevant resources.

Previous studies \cite{caragea2014ecir,wu2017supervised} used only metadata in the header of scholarly articles for paper entity linking. In reality, the quality of a header is not always that good. Hence, we investigated leveraging both header and citation information to match paper entities between two digital library datasets when the target corpus contains noisy data. We proposed an approach that integrates header and citation information for paper entity matching. Compared with the IR-based method, header matching model improves precision by at least $13\%$ and F1 by about $3\%$ for papers with low quality titles. The integrated model with header and citation information achieves an F1 as high as $0.992$ and precision as high as $0.984$.  We show that \csx\ has a huge overlap with WoS, DBLP, and Medline, which can be used for metadata correction, and that there are still a large number of scientific documents to be crawled and indexed.  The framework developed can be used to match records between any bibliographic databases with or without citations. The idea of combining ML and IR is in general applicable to many information retrieval and data linking problems. We will apply this framework to clean the \csx\ metadata. This will generate high quality large-scale datasets that can enable development and implementation of many graph-based AI applications. 

The software implementation of this framework is available on GitHub at \\
{\small \url{https://github.com/SeerLabs/entity-matching}.}
\subsection{Acknowledgements}
We gratefully acknowledge partial support from the National Science Foundation.

\bibliographystyle{aaai}
\end{document}